\documentclass[preprintnumbers,prd,twocolumn,showpacs,floatfix,preprintnumbers,amsmath,amssymb,nofootinbib,superscriptaddress]{revtex4}
\usepackage{graphicx}
\usepackage{epsfig}
\usepackage{bm}
\usepackage{amsfonts}
\usepackage{color}
\usepackage{subfigure}
\usepackage{amsmath}
\usepackage{amssymb}
\usepackage{graphicx}
\usepackage{makecell}
\usepackage{float}

\usepackage{amsmath,amssymb}

\begin{document}

\title{A model independent parametrization of the late time cosmic acceleration: constraints on the parameters from recent observations}

\author{Bikash R. Dinda}
\email{bikashd18@gmail.com}
\affiliation{Department of Theoretical Physics, Tata Institute of Fundamental Research,\\Dr Homi Bhabha Road, Navy Nagar, Colaba, Mumbai-400005, India.}

\begin{abstract}
In this work, we have considered a model independent approach to study the nature of the late time cosmic acceleration. We have used the Pade approximation to parametrize the comoving distance. Consequently, from this comoving distance, we derive a parameterization for the Hubble parameter. Our parametrization is completely analytic and valid for late-time and matter dominated eras only. This parametrization possesses sub-percentage accuracy compared to any arbitrary cosmological model or parametrization up to matter dominated era. Using this parametrization, we put constraints on the parameters from recent low redshift cosmological observations including Planck 2018 distance priors. Our results show that the $\Lambda$CDM model is 1$\sigma$ to 2$\sigma$ away for lower redshifts. We find that the phantom crossing is allowed by all the combinations of dataset considered. We also find that the dynamical dark energy models are preferable at lower redshifts. Our study also shows that, at lower redshifts ($z<0.5$), phantom models are allowed at almost 1$\sigma$ confidence level.
\end{abstract}


\maketitle
\date{\today}

\section{Introduction}

In last two decades, many cosmological observations like Supernova Type-Ia observations \citep{SnIa1,SnIa2}, cosmic microwave background observations \citep{CMB1,CMB2,Planck}, baryon acoustic oscillations measurements \citep{BAO1,BAO2} have strongly confirmed that the Universe is undergoing through an accelerated era at late times. This acceleration can be explained by introducing an exotic matter, called dark energy \citep{ThQ7,ThQ8,Tracker1,Tracker2,ThQ1,ThQ2,ThQ3,ThQ4,ThQ5,ThQ6} (which has large negative pressure) or by the modification of gravity \citep{mog1,mog2,mog3,mog4,mog5}. In literature, there are many dark energy models like quintessence and k-essence \citep{ThQ7,ThQ8}. Also there are many modified gravity models like f(R) gravity \citep{fR1,fR2,fR3}, Galileon gravity \citep{galBasic1,galBasic2,galBasic3} and Horndeski gravity \citep{Hrnski}.

One of the simplest dark energy models is the $\Lambda$CDM model. However, the $ \Lambda $CDM model has some theoretical problems like the fine-tuning and cosmic coincidence problems \citep{fntn}. Besides these theoretical problems, recently some observations \citep{inctny1,inctny2,inctny3,inctny4} have suggested that the $\Lambda$CDM model is not exactly the best fit model to the recent low redshift cosmological data and dynamical dark energy is preferred. These results motivate us to go beyond the $ \Lambda $CDM model and preferably for the dynamical dark energy models.

Although there are many dynamical dark energy or modified gravity models to explain the late time cosmic acceleration, it is still necessary to do the detailed analysis of the behavior of the late time acceleration from cosmological observations. For this reason, proper parametrization of the late time cosmic acceleration is needed, especially, the model independent parametrization. For example, if most of the observations suggest phantom crossing for the equation of state of the dark energy, use of quintessence model to study the nature of the late time cosmic acceleration will provide us incomplete information. For this reason, in literature, there are some parametric approach to the equation of state of the dark energy like CPL parametrization \citep{CPL1,CPL2}, BA parametrization \citep{BA} and GCG parametrization \citep{GCG}. Among these, CPL parametrization is the most popular parametrization in the literature. It includes both the phantom and non-phantom behavior of the equation of state of dark energy.

Most of the above-mentioned parametrizations based on the Taylor series expansion. However, Parametrizations based on Taylor series expansion can only give us accurate results if the argument (for example, $(1-a)$ in the CPL parameterization with $a$ being the scale factor) of a particular function is much less than unity. In this regards, the Pade approximation \citep{Pade1,Pade2} to any cosmological quantity can give us better results compared to the Taylor series expansion (with the same number of parameters involved). In literature, Pade approximation has already been used like in \citep{Pade2,Padeapply1,Padeapply2} for the equation of state of dark energy, in \citep{Padeapply3} for the energy density of the dark energy, in \citep{Padeapply4,Padeapply5} for luminosity distance, etc.

Here, we also use Pade approximation to parametrize the comoving distance and consequently to the Hubble parameter. Most of the above-mentioned parametrizations (using Pade approximation) give better accuracy as long as redshift is not very high (where the Pade approximation uses redshift as the argument). However, here, we provide a simple parametrization (based on Pade approximation) which can provide better accuracy at higher redshifts too by using some cosmological information. We discuss this in the next section.

\section{Parametrization to the comoving distance and Hubble parameter}

Here, our final aim is to parametrize the comoving distance and from it finally to parametrize the Hubble parameter. To do this, we do not directly parametrize it. The reason for this will be discussed soon. The comoving distance (line of sight), $\chi$ is defined as

\begin{equation}
\chi (z) = d_{H} \int_{0}^{z} \frac{dz'}{E(z')},
\label{eq:dcov}
\end{equation}

\noindent
where $z$ is the redshift. $E$ is the normalized Hubble parameter, defined as $E(z)=H(z)/H_{0}$. $H$ is the Hubble parameter and $H_{0}$ is the present day ($z=0$) Hubble parameter. Here $d_{H}=1/H_{0}$. The derivative of the Eq.~\eqref{eq:dcov} gives us a relation given by

\begin{equation}
\frac{1}{E(z)} = \frac{1}{d_{H}} \left[ \dfrac{d\chi(z)}{dz} \right],
\label{eq:EinvdchiNpre}
\end{equation}

\noindent
Now, we define two dimensionless variables given by

\begin{eqnarray}
\tilde{E}(z) &=& \frac{E(z)}{\sqrt{\Omega_{m0}}}, \nonumber\\
\chi^{N}(z) &=& \frac{\sqrt{\Omega_{m0}}}{d_{H}} \chi(z),
\label{eq:dmnlsvrbl}
\end{eqnarray}

\noindent
where $\Omega_{m0}$ is the matter energy density parameter at present. Using the above equation, Eq.~\eqref{eq:EinvdchiNpre} can be rewritten in dimensionless variables given by

\begin{equation}
\frac{1}{\tilde{E}(z)} = \dfrac{d\chi^{N}(z)}{dz},
\label{eq:EinvdchiN}
\end{equation}

\noindent
If the Universe is dominated by matter only, then $\tilde{E}(z)$ becomes (denoted by $\tilde{E}_{\text{matter}}$)

\begin{equation}
\tilde{E}_{\text{matter}}(z) = (1+z)^{3/2}.
\label{eq:Ematter}
\end{equation}

\noindent
Putting Eq.~\eqref{eq:Ematter} into Eq.~\eqref{eq:dcov}, we get normalized comoving distance (denoted by $\chi^{N}_{\text{matter}}$) for matter dominated Universe given by

\begin{equation}
\chi^{N}_{\text{matter}}(z) = 2 \left[ \frac{\sqrt{1+z}-1}{\sqrt{1+z}} \right]+I_{c}.
\label{eq:chiNmatter}
\end{equation}

\noindent
Here we are interested in the matter dominated and late time eras only. Now, let say, due to the presence of dark energy or due to the modification of gravity, total normalized comoving distance differs from the one in Eq.~\eqref{eq:chiNmatter}. Let say, the correction is denoted by $\chi^{N}_{c}$. So, we have

\begin{equation}
\chi^{N}(z)=\chi^{N}_{\text{matter}}(z)+\chi^{N}_{c}(z).
\label{eq:chiNtot}
\end{equation}

\noindent
Note that, we have not considered $\chi^{N}_{\text{matter}}$ to be zero at $z=0$, because it is not necessary as long as $\chi^{N}(z=0)=0$. That is why we have introduced an integral constant, $I_{c}$ in Eq.~\eqref{eq:chiNmatter}.

At this stage, one can parametrize the extra correction, $\chi^{N}_{c}(z)$ both in a model dependent and independent ways. One of the model independent way could be to expand it around the present time ($z=0$) using Taylor series up to a particular order. The coefficients of this Taylor series can be the model-independent parameters to be constrained. However, with Taylor series expansion, one has to consider a large number of the order to get better accuracy. This corresponds to a large number of parameters. Also, for $z\gg1$, the accuracy decreases rapidly. To avoid these issues, here, we consider Pade expansion to get an approximate series for $\chi^{N}_{c}(z)$. This can reduce the error compared to the Taylor series expansion. Now, we can expand $\chi^{N}_{c}(z)$ with the Pade approximation with (m,n) order (denoted by $P^{m}_{n}$) given by

\begin{equation}
\chi^{N}_{c}(z) \approx P^{m}_{n} \left[\chi^{N}_{c}(z)\right] = \dfrac{\sum_{i=0}^{m} P_{i} z^{i}}{\sum_{j=0}^{n} Q_{j} z^{j}},
\label{eq:PadechiNc}
\end{equation}

\noindent
where $P_{i}$'s and $Q_{j}$'s are the parameters of the Pade expansion. We can always put $Q_{0}=1$ because it will normalize the series both in numerator and denominator. So, there are $(1+m+n)$ number of independent parameters in a $P^{m}_{n}$ order Pade expansion series. Although there is a finite number of independent parameters, the Pade series is infinite (however, note that the higher-order coefficients are not independent in this infinite series). Because of this infinite series expansion, $P^{m}_{n}$ order Pade expansion series has better accuracy compared to a finite Taylor series expansion up to $(m+n+1)$th order (although the number of independent parameters is same as $1+m+n$ for both the cases).

\noindent
So far we have not put any cosmological information about the extra correction. So, our Pade approximation should have knowledge about this extra term mentioned below:

\begin{itemize}
\item
At present, $\chi^{N}(z=0)$ should be zero. This gives

\begin{equation}
I_{c} = -P_{0}.
\label{eq:IcwrtP0}
\end{equation}
\end{itemize}

\noindent
So, total normalized comoving distance immediately becomes

\begin{equation}
\chi^{N}(z)=2 \left[ \frac{\sqrt{1+z}-1}{\sqrt{1+z}} \right]-P_{0}+\dfrac{P_{0}+\sum_{i=1}^{m} P_{i} z^{i}}{1+\sum_{j=1}^{n} Q_{j} z^{j}}.
\label{eq:chiNtot2}
\end{equation}

\noindent
Using Eq.~\eqref{eq:EinvdchiN}, we can now calculate the inverse of $\tilde{E}$ given by

\begin{equation}
\frac{1}{\tilde{E}_{\text{tot}}(z)}=(1+z)^{-3/2}+A(z),
\label{eq:Einvtot}
\end{equation}

\noindent
where $A(z)$ is given by

\begin{eqnarray}
A(z)&&=\frac{\left[A_{1}(z)-A_{2}(z)\right]}{\left(1+\sum_{j=1}^{n} Q_{j} z^{j}\right)^{2}}, \nonumber\\
&& A_{1}(z) = \left(\sum_{i=1}^{m} i P_{i} z^{i-1}\right)\left(1+\sum_{k=1}^{n} Q_{k} z^{k}\right), \nonumber\\
&& A_{2}(z) = \left(P_{0}+\sum_{i=1}^{m} P_{i} z^{i}\right)\left(\sum_{k=1}^{n} k Q_{k} z^{k-1}\right).
\label{eq:Aofz}
\end{eqnarray}

\noindent
This $A(z)$ is basically the correction to normalized inverse Hubble parameter due to dark energy or modification to the gravity. Now, we use other cosmological information given by

\begin{itemize}
\item
At present, $E_{\text{tot}}(z=0)$ should be unity i.e. $\tilde{E}(z=0)=1/\sqrt{\Omega_{m0}}$. This gives

\begin{equation}
P_{1}=P_{0}Q_{1}+\sqrt{\Omega_{m0}}-1.
\label{eq:P1wrtP0Q1}
\end{equation}

\item
At matter dominated era, $A(z)$ should be negligible compared to $1/\tilde{E}_{\text{matter}}(z)$ i.e. at higher redshifts, the extra effect should be negligible. This is possible if the highest order in $z$ in the denominator is larger than the numerator in $A(z)$ compared to the matter term (and since Pade parameters should be of the same order of magnitude). Using this fact, we can put one restriction to the arbitrary (m,n) order of the $P^{m}_{n}$ series given by

\begin{equation}
n \geq m+1 \hspace{1 cm} \text{with} \hspace{1 cm} m \geq 0.
\label{eq:nmrestriction}
\end{equation}

\end{itemize}

Note that from $(1+m+n)$ number of parameters one is now fixed ($P_{1}$ from Eq.~\eqref{eq:P1wrtP0Q1}). So, finally, the number of independent (Pade) parameters is $m+n$. So, we obtain the final expression for the comoving distance given by

\begin{eqnarray}
&& \chi(z)=\frac{d_{H}}{\sqrt{\Omega _{m0}}}\Big{[}2 \left[ \frac{\sqrt{1+z}-1}{\sqrt{1+z}} \right]-P_{0} \nonumber\\
&& +\dfrac{P_{0}+\left(P_{0}Q_{1}+\sqrt{\Omega_{m0}}-1\right)z+\sum_{i=2}^{m} P_{i} z^{i}}{1+\sum_{j=1}^{n(n \geq m+1)} Q_{j} z^{j}}\Big{]}.
\label{eq:chiFinal}
\end{eqnarray}

\noindent
Consequently, the Hubble parameter becomes

\begin{eqnarray}
H(z) = \frac{H_{0} \sqrt{\Omega_{m0}}}{(1+z)^{-3/2}+A(z)}.
\label{eq:PadeHFinal}
\end{eqnarray}

\noindent
Here, independent parameters are $P_{0},P_{2},...,P_{m};Q_{1},Q_{2},...,Q_{n}$.

\subsection{Singularity issue}
Note that there is a singularity issue in the Eqs.~\eqref{eq:chiFinal} and~\eqref{eq:PadeHFinal} (or consequently in other equations) as in any Pade series. In the broad range of the parameter space, for some particular values of the parameters, the denominator can be zero i.e.

\begin{equation}
1+\sum_{j=1}^{n} Q_{j} z^{j} = B(z) \hspace{0.2 cm}(\text{say}) = 0 \rightarrow \text{singularity}.
\label{eq:singularity}
\end{equation}

\noindent
So, when the denominator becomes zero, the Pade series diverges. We can overcome this singularity by assuming that any observable should be monotonic functions of $z$. So, to avoid singularity issue, we make a correction given by

\begin{equation}
\chi^{N}_{c}(z) = \begin{cases}
\chi^{N}_{c}(z) \hspace{2.3 cm} \text{for} \hspace{0.2 cm} B(z)\neq0 \\ \frac{\chi^{N}_{c}(z-\Delta)+\chi^{N}_{c}(z+\Delta)}{2} \hspace{0.5 cm} \text{for} \hspace{0.2 cm} B(z)=0
\end{cases},
\label{eq:chiNccorrected}
\end{equation}

\noindent
where $\Delta$ is introduced to represent a small number. We can safely take

\begin{equation}
\Delta = 10^{-3}.
\label{eq:Delta}
\end{equation}

\noindent
Note that one can take any other values of $\Delta$ as long as $\Delta<<1$. One can check that the error arises due to this correction is insignificant. So, this correction will not lead to any major changes in our formalism. Note that, in most of the cases, the singularity will not arise, unless we allow some arbitrary values of the parameters far from the standard cosmological values.

\section{Example: $m=1$ and $n=2$ case:}

For $m=1$ and $n=2$ case, the comoving distance (from Eq.~\eqref{eq:chiFinal}) becomes

\begin{eqnarray}
&& \chi(z)=\frac{d_{H}}{\sqrt{\Omega _{m0}}}\Big{[}2 \left[ \frac{\sqrt{1+z}-1}{\sqrt{1+z}} \right]-P_{0} \nonumber\\
&& +\dfrac{P_{0}+\left(P_{0}Q_{1}+\sqrt{\Omega_{m0}}-1\right)z}{1+Q_{1}z+Q_{2}z^{2}}\Big{]}.
\label{eq:chiFinal12}
\end{eqnarray}

\noindent
Consequently, the Hubble parameter (from Eq.~\eqref{eq:PadeHFinal}) becomes

\begin{eqnarray}
H(z) = \frac{H_{0} \sqrt{\Omega_{m0}}}{(1+z)^{-3/2}+A(z)},
\label{eq:PadeHFinal12}
\end{eqnarray}

\noindent
with $A(z)$ is given by (from Eq.~\eqref{eq:Aofz})

\begin{eqnarray}
A(z)&=&\frac{\tilde{A}(z)}{\left(1+Q_{1}z+Q_{2}z^{2}\right)^{2}}, \nonumber\\
&& \tilde{A}(z) = (\sqrt{\Omega_{m0}}-1)-2 P_0 Q_2 z \nonumber\\
&& -(P_{0}Q_{1} +\sqrt{\Omega_{m0}}-1) Q_2 z^2.
\label{eq:Aofz12}
\end{eqnarray}

\noindent
respectively. Eqs.~\eqref{eq:chiFinal12} and~\eqref{eq:PadeHFinal12} are the main result for the parametrization of the comoving distance and the Hubble parameter for the $m=1$, $n=2$ case. And our independent parameters are $P_{0}$, $Q_{1}$, and $Q_{2}$ with the cosmological parameters $\Omega_{m0}$ and $H_{0}$.

\section{Advantage of our method over Taylor series and normal Pade series}
\label{sec-fiducialfitting}

The Hubble parameter derived in Eq.~\eqref{eq:PadeHFinal} (or in Eq.~\eqref{eq:PadeHFinal12}) is now useful to use for higher redshifts as compared to direct Taylor series expansion or normal Pade series expansion to the Hubble parameter. To show this exclusively, we first consider a fiducial model. Then we compare our result to this fiducial model. Also, we compare the Taylor series and normal Pade series results in this fiducial model. In this way, we can compare our result to the results from the Taylor series and Normal Pade series results. What we do here is that we define a chi-square considering the fiducial model as the base. Note that, here no real data is involved. The real data analysis is presented in the next section. Now, the chi-square can be written as (considering normalized Hubble parameter is the quantity to be compared for different models with the fiducial model)

\begin{equation}
\text{chi}^{2} = \sum_{z=0}^{z_{max}} \frac{\left( E(z)-E_{fid}(z) \right)^{2}}{\left( \Delta E_{fid}(z) \right)^{2}},
\label{eq:chisqrfid}
\end{equation}

\noindent
where $E(z)$ is the normalized Hubble parameter for any model (be it our expansion series or Taylor series or normal Pade series). The entity with the subscript 'fid' corresponds to the fiducial model. Since there is no real observation involved here, we don't know the error ($\Delta E_{fid}(z)$ in Eq.~\eqref{eq:chisqrfid}) in the fiducial result. However, the error is not important here, because we are interested only in the best fit values of the parameters from this chi-square minimization. So, we consider an arbitrary error and it is $\Delta E_{fid}(z) = E_{fid}(z)/10$ (i.e. the error is $10\%$ of the main value at each redshift respectively). One can check that the best fit parameter values are not sensitive to the values of the error. Here, the summation is over redshift interval. We consider it to be $\Delta z = 0.01$. For the upper limit of redshift, we consider $z_{max}=6$.

\subsection{Fiducial model}

Here, we have considered CPL parametrization to be the fiducial model. In the CPL parametrization, the square of the normalized Hubble parameter is given by \citep{CPL1,CPL2}

\begin{eqnarray}
E^{2}_{fid}(z) &&= E^{2}_{CPL}(z) = \Omega_{m0}(1+z)^{3} \nonumber\\
&& +(1-\Omega_{m0})(1+z)^{3(1+w_{0}+w_{a})} e^{-\frac{3w_{a}z}{(1+z)}},
\label{eq:fidE}
\end{eqnarray}

\noindent
where $w_{0}$ and $w_{a}$ are two CPL parameters describing the evolution of the equation of state of dark energy given by $w(z)=w_{0}+w_{a}\frac{z}{1+z}$. In this CPL parametrization, we consider three fiducial models, mentioned below

\begin{itemize}
\item
fiducial model 1 (denoted by fid1): $w_{0}=-1$, $w_{a}=0$ ($\Lambda$CDM),
\item
fiducial model 2 (denoted by fid2): $w_{0}=-1.2$, $w_{a}=-0.2$ (phantom DE),
\item
fiducial model 3 (denoted by fid3): $w_{0}=-0.8$, $w_{a}=0.2$ (non-phantom DE).
\end{itemize}

\noindent
Now, considering each of these fiducial models, we perform chi-square minimization to obtain the best fit parameter values in our model. And the values are listed in Table ~\ref{table:tbl1}.

\begin{table}[tbp]
\begin{center}
\begin{tabular}{ | m{14em} | m{2cm}| } 
\hline
fiducial model& parameter values in our model \\ 
\hline
fid1: ($\Lambda$CDM)& $P_{0}=0.1867$ $Q_{1}=2.053$ $Q_{2}=1.87$ \\ 
\hline
fid2: ($w_{0}=-1.2$, $w_{a}=-0.2$) & $P_{0}=0.155$ $Q_{1}=2.366$ $Q_{2}=3.087$ \\ 
\hline
fid3: ($w_{0}=-0.8$, $w_{a}=0.2$) & $P_{0}=0.24$ $Q_{1}=1.7568$ $Q_{2}=1.0$ \\ 
\hline
\end{tabular}
\end{center}
    \caption{Parameter values of our model from chi-square minimization using Eq.~\eqref{eq:chisqrfid} (and the normalized Hubble parameter from Eq.~\eqref{eq:PadeHFinal12}).} 
    \label{table:tbl1}
\end{table}

\noindent
Before, presenting results from this best fit values, let us first briefly discuss Taylor series expansion and normal Pade series expansion direct to the Hubble parameter.

\subsection{Taylor series}

The Taylor series expansion of the normalized Hubble parameter is given by (up to fourth order)

\begin{equation}
E(z)(\text{Taylor series}) = 1+a_{1}z+a_{2}z^{2}+a_{3}z^{3}+a_{4}z^{4},
\label{eq:TaylorE}
\end{equation}

\noindent
where, $a_{1}$, $a_{2}$, $a_{3}$ and $a_{4}$ are four parameters. Doing similar chi square minimization, we get best fit values of these parameters. These are listed in Table ~\ref{table:tbl2}.

\begin{table}[tbp]
\begin{center}
\begin{tabular}{ | m{14em} | m{2cm}| } 
\hline
fiducial model& parameter values in Taylor series \\ 
\hline
fid1: ($\Lambda$CDM)& $a_{1}=0.47$ $a_{2}=0.33$ $a_{3}=-0.04$ $a_{4}=0.003$ \\ 
\hline
fid2: ($w_{0}=-1.2$, $w_{a}=-0.2$) & $a_{1}=0.285$ $a_{2}=0.45$ $a_{3}=-0.07$ $a_{4}=0.005$ \\ 
\hline
fid3: ($w_{0}=-0.8$, $w_{a}=0.2$) & $a_{1}=0.68$ $a_{2}=0.235$ $a_{3}=-0.02$ $a_{4}=0.001$ \\ 
\hline
\end{tabular}
\end{center}
    \caption{Parameter values in the Taylor series from chi-square minimization using Eq.~\eqref{eq:chisqrfid} (and the normalized Hubble parameter from Eq.~\eqref{eq:TaylorE}).} 
    \label{table:tbl2}
\end{table}

\subsection{Normal Pade series}

The (2,2) order Pade series expansion direct to the normalized Hubble parameter is given by (for details see \citep{Capozziello:2018jya})

\begin{equation}
E(z)(\text{normal Pade series}) = \frac{1+p_{1}z+p_{2}z^{2}}{1+q_{1}z+q_{2}z^{2}},
\label{eq:normalPadeE}
\end{equation}

\noindent
where, $p_{1}$, $p_{2}$, $q_{1}$ and $q_{2}$ are four Pade parameters. Here, we have denoted the Pade parameters by small letters to distinguish these from our parameters. Doing similar chi-square minimization, we get best-fit values of these parameters. These are listed in Table ~\ref{table:tbl3}.

\begin{table}[tbp]
\begin{center}
\begin{tabular}{ | m{14em} | m{2cm}| } 
\hline
fiducial model& parameter values in normal Pade series \\ 
\hline
fid1: ($\Lambda$CDM)& $p_{1}=0.699$ $p_{2}=0.50$ $q_{1}=0.26$ $q_{2}=-0.007$ \\ 
\hline
fid2: ($w_{0}=-1.2$, $w_{a}=-0.2$) & $p_{1}=0.63$ $p_{2}=0.73$ $q_{1}=0.43$ $q_{2}=-0.01$ \\ 
\hline
fid3: ($w_{0}=-0.8$, $w_{a}=0.2$) & $p_{1}=0.848$ $p_{2}=0.38$ $q_{1}=0.18$ $q_{2}=-0.005$ \\ 
\hline
\end{tabular}
\end{center}
    \caption{Parameter values in the normal Pade series from chi-square minimization using Eq.~\eqref{eq:chisqrfid} (and the normalized Hubble parameter from Eq.~\eqref{eq:normalPadeE}).} 
    \label{table:tbl3}
\end{table}

\noindent
The reason to consider fourth-order Taylor and Pade series is that in these series, the $\Omega_{m0}$ parameter is not exclusive. And in our series expansion, it is an external parameter. So, to have a fair comparison, we have considered fourth-order series expansion both in Taylor and Pade series, although we have fixed $\Omega_{m0}=0.3$ here (in this section only).

\begin{figure}[tbp]
\centering
\includegraphics[width=.45\textwidth]{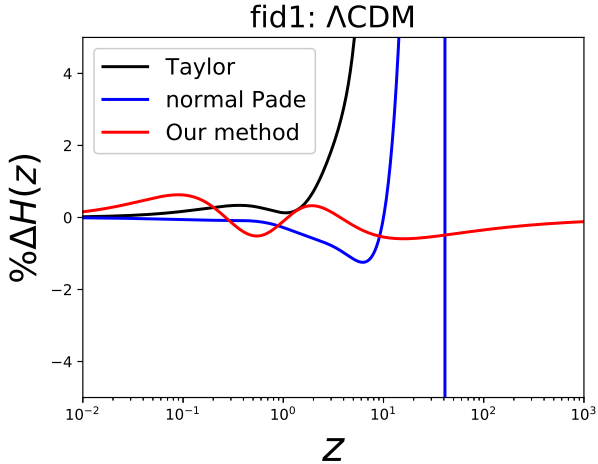}
\caption{\label{fig:Hcmp1} Percentage deviations in the Hubble parameter (using Eq.~\eqref{eq:prcntgdltH}) in our series vs. Taylor series vs. normal Pade series compared to the fiducial model 1 (denoted by fid1 which is the $\Lambda$CDM model). Black, blue and red lines correspond to the results from the Taylor series, normal Pade series and our series respectively.}
\end{figure}

\begin{figure}[tbp]
\centering
\includegraphics[width=.45\textwidth]{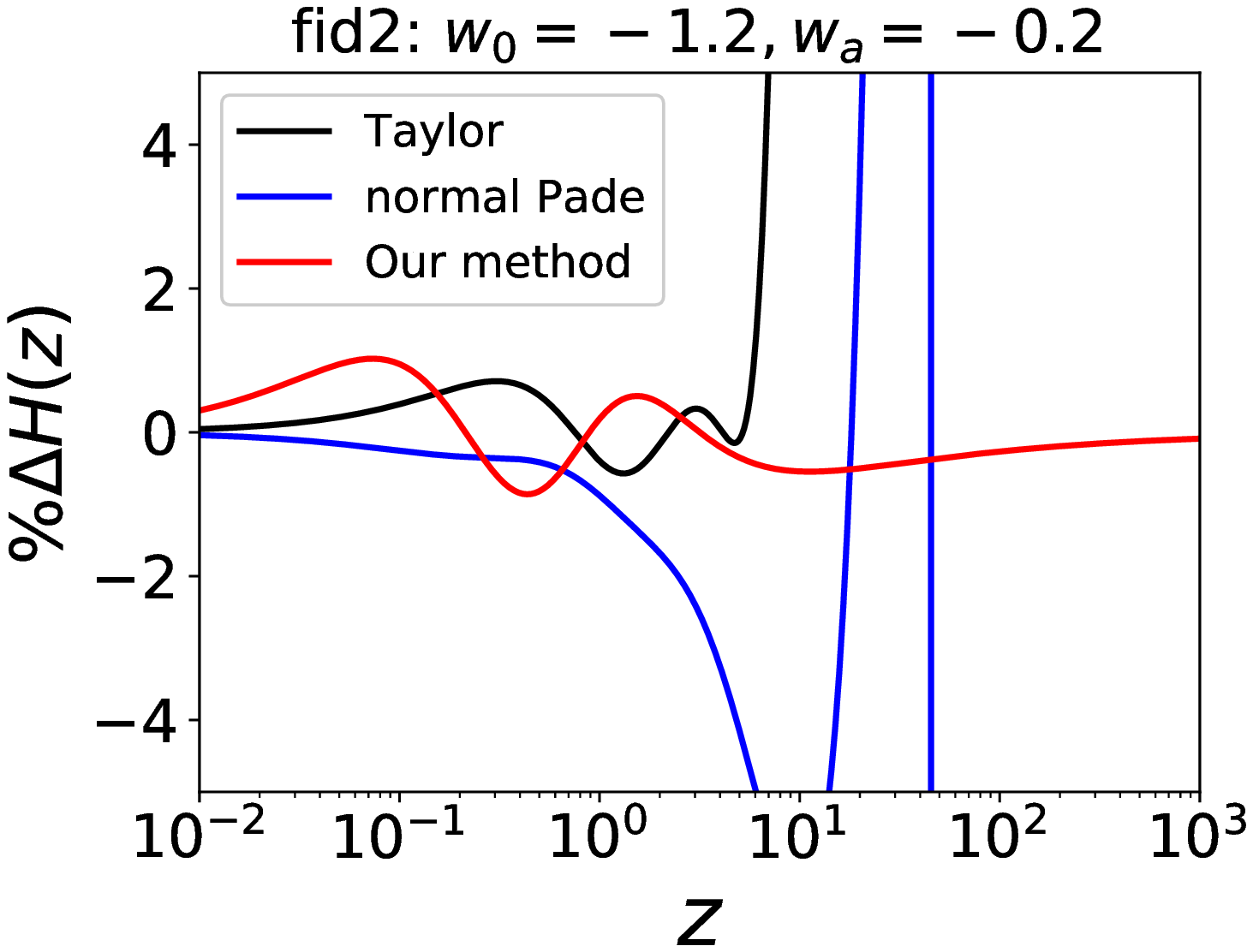}
\caption{\label{fig:Hcmp2} Percentage deviations in the Hubble parameter (using Eq.~\eqref{eq:prcntgdltH}) in our series vs. Taylor series vs. normal Pade series compared to the fiducial model 2 (denoted by fid2 which is $w_{0}=-1.2$, $w_{a}=-0.2$ model in CPL parametrization). Black, blue and red lines correspond to the results from the Taylor series, normal Pade series and our series respectively.}
\end{figure}

\begin{figure}[tbp]
\centering
\includegraphics[width=.45\textwidth]{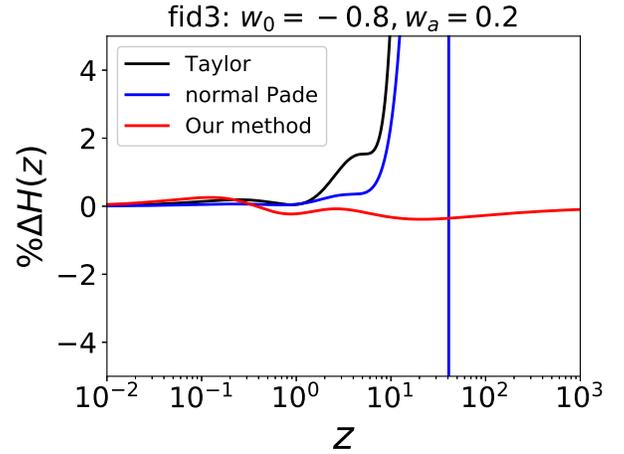}
\caption{\label{fig:Hcmp3} Percentage deviations in the Hubble parameter (using Eq.~\eqref{eq:prcntgdltH}) in our series vs. Taylor series vs. normal Pade series compared to the fiducial model 3 (denoted by fid3 which is $w_{0}=-0.8$, $w_{a}=0.2$ model in CPL parametrization). Black, blue and red lines correspond to the results from the Taylor series, normal Pade series and our series respectively.}
\end{figure}

Using the above best fit parameter values from Table ~\ref{table:tbl1},~\ref{table:tbl2} and~\ref{table:tbl3}, we now compare the evolution of the Hubble parameter with each fiducial model respectively. The comparison plots are shown in Figs.~\ref{fig:Hcmp1},~\ref{fig:Hcmp2} and~\ref{fig:Hcmp3} for fid1, fid2 and fid3 models respectively. In these figures, $\%\Delta H$ denotes the percentage deviation of the Hubble parameter in three different models from a particular fiducial model. It is defined as

\begin{equation}
\%\Delta H(z)= \frac{H(z)-H_{\text{fid}}(z)}{H_{\text{fid}}(z)} \times 100.
\label{eq:prcntgdltH}
\end{equation}

\noindent
In the above equation, for a particular fiducial model, we compute $H(z)$ in a particular model (be it our series or Taylor series or normal Pade series) with the best fit parameters (mentioned in Tables ~\ref{table:tbl1},~\ref{table:tbl2} and~\ref{table:tbl3}). For example, in Fig.~\ref{fig:Hcmp1}, the red line corresponds to the percentage deviation in Hubble parameter in our series expansion compare to the fid1 model (i.e. $\Lambda$CDM). First, we compute $H(z)$ from Eq.~\eqref{eq:PadeHFinal12} by putting the parameter values, mentioned in first row (second column) in Table ~\ref{table:tbl1}. Then we put this $H(z)$ in Eq.~\eqref{eq:prcntgdltH} to get $\%\Delta H(z)$ with $H_{\text{fid}}(z)$ corresponding to $H(z)$ for $\Lambda$CDM. Similarly, black and blue lines are for Taylor series and normal Pade series respectively for fid1 in in Fig.~\ref{fig:Hcmp1}. In similar fashion, Figs.~\ref{fig:Hcmp2} and~\ref{fig:Hcmp3} represent percentage deviations from fid2 and fid3 respectively.

So, Figs.~\ref{fig:Hcmp1},~\ref{fig:Hcmp2} and~\ref{fig:Hcmp3} show that how good our parametrization can fit to an arbitrary fiducial model. In all these figures, we have fixed $h=0.7$ (for this section only), where $h$ is defined as $H_{0}=100hkm/s/Mpc$. We can see that the errors are up to $1\%$ level for a large redshift range ($0\leq z \leq 1000$) in our method. For Taylor series, the errors are at percentage level up to $z=1$. After that, the errors increase rapidly. For normal Pade series, the errors are at the percentage level up to $z=5-7$. After that, the errors increase drastically. So, our series expansion gives better result compared to the Taylor series or normal Pade series.

Note that, if one considers smaller values of $z_{max}$ for the chi-square minimization, the best fit values of the parameters change accordingly and the errors decrease with decreasing $z_{max}$.

Here, we have compared the results in the Hubble parameter (or in the normalized Hubble parameter). The results will be similar if we consider the comoving distance or any other background quantity.

So, the conclusion here is that any Taylor series or Pade approximation, which directly parametrize Hubble parameter or comoving distance (or luminosity distance) can not fit any arbitrary model at sub percentage level for large redshift ranges. Splitting up the comoving distance and proper utilization of cosmological information give us a better parametrization. This is a big advantage to use our parametrization. Other advantages of this parametrization are listed in the conclusion section. Note that, the errors will decrease further if one considers higher values of $m$ and $n$.

\section{Constraints on $P_{0}$, $Q_{1}$ and $Q_{2}$ from recent observations}
\label{sec-dataanalysis}

\begin{figure}[tbp]
\centering
\includegraphics[width=.5\textwidth]{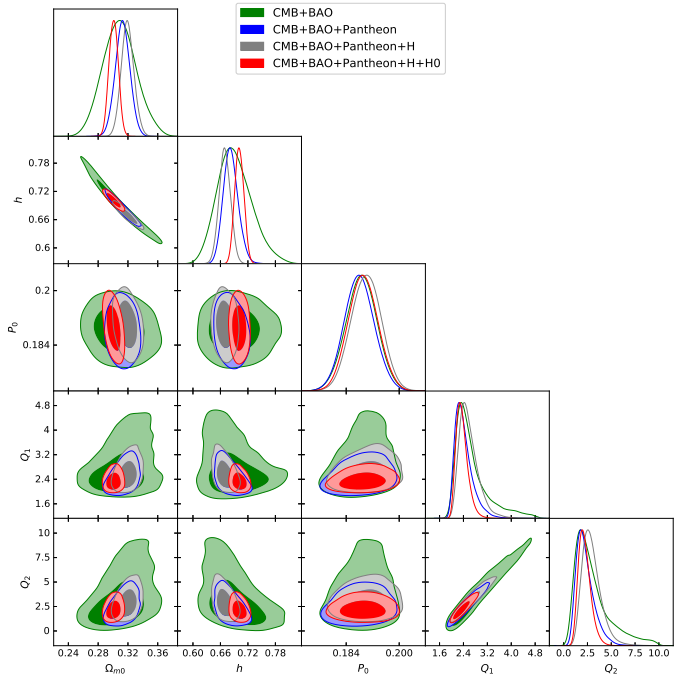}
\caption{\label{fig:triangleplot} Triangle plot for the parameters (parameters in our series and the cosmological parameters) using recent cosmological observations mentioned in the text (in section-\ref{sec-dataanalysis}).}
\end{figure}

In Figure~\ref{fig:triangleplot}, we have shown the triangle plot i.e. 1$\sigma$ and 2$\sigma$ confidence contours of all the pairs of the parameters and likelihood of every parameter for different combinations of the observations. The data used here are given below:

\begin{itemize}
\item
(1) The BAO measurements from different surveys. These surveys are 6DF survey \citep{6dF}, SDSS survey for galaxy sample (MGS) \citep{MGS} and eBOSS quasar clustering \citep{eBOSSq}, and the Lyman-$\alpha$ forest sample \citep{Lyalpha}.
\item
(2) Measurement of angular diameter distances from water megamasers under the Megamaser Cosmology Project \citep{masers1,masers2,masers3}. (1) and (2) data together, we denote this as 'BAO' \citep{Evslin:2017qdn,Capozziello:2018jya,Dutta:2018vmq,Lonappan:2017lzt}.
\item
(3) Recent Pantheon data for SNIa observation \citep{Panth1,Panth2}. We denote this data as 'Pantheon'.
\item
(4) The OHD data for the Hubble parameter compiled in Pinho et. al. \citep{OHD}. We denote this data as 'H'.
\item
(5) The latest measurement of $H_{0}$ by Riess et. al. (2019) \citep{RiessH0}. The value is given by $H_{0}=74.03 \pm 1.42 km/s/Mpc$. We denote this data as '$H_{0}$'.
\item
(6) We also use Planck 2018 results (TT,TE,EE+lowl+lowE for the base $\Lambda$CDM model) as prior \citep{Planck18,Planck18dist}. We denote this as 'CMB'.
\end{itemize}

Using these data sets, in Figure~\ref{fig:triangleplot}, we have chosen four combinations of dataset given by 'CMB+BAO' (shaded with green color), 'CMB+BAO+Pantheon' (shaded with blue color), 'CMB+BAO+Pantheon+H' (shaded with grey color) and 'CMB+BAO+Pantheon+H+$H_{0}$' (shaded with red color). For every combination (i.e. for every color), the dark and light-shaded regions are for 1$\sigma$ and 2$\sigma$ confidence contours respectively. We use these same color combinations for the next plots also.

\begin{table}[tbp]
    \begin{center}
\begin{tabular} {l l l l c c c c}

\hline 
{         } & \text{C+B}  & \text{C+B+P}   & \text{C+B+P+H}   & \text{All}    \\
\hline                                                     \\

{\boldmath$\Omega_{m0}    $} & $0.309^{+0.046}_{-0.043}$ & $0.313^{+0.019}_{-0.019}$ & $0.320^{+0.017}_{-0.016}   $ & $0.301^{+0.012}_{-0.012}   $ \\ \\

{\boldmath$h              $} & $0.691^{+0.076}_{-0.070}$  & $0.683^{+0.034}_{-0.030}   $ & $0.670^{+0.024}_{-0.024}   $ & $0.701^{+0.019}_{-0.019}   $ \\ \\

{\boldmath$P_{0}          $} & $0.1886^{+0.0092}_{-0.0092}   $ & $0.1880^{+0.0091}_{-0.0088}   $ & $0.1900^{+0.0087}_{-0.0088}$ & $0.1891^{+0.0089}_{-0.0086}$ \\ \\

{\boldmath$Q_{1}          $} & $2.67^{+1.3}_{-0.77}      $ & $2.41^{+0.63}_{-0.48}      $ & $2.58^{+0.66}_{-0.55}      $ & $2.35^{+0.41}_{-0.35}      $ \\ \\

{\boldmath$Q_{2}          $} & $3.0^{+4.3}_{-2.6}         $ & $2.3^{+2.0}_{-1.6}       $ & $2.9^{+2.0}_{-1.7}         $ & $2.2^{+1.3}_{-1.2}         $ \\ \\
\hline
\end{tabular}
    \end{center}
\caption{Marginalized $95\%$ 1D confidence intervals of the parameters in our series and the cosmological parameters from the different combinations of data mentioned in section-\ref{sec-dataanalysis}.}
\label{table:tbl4}
\end{table}

Table ~\ref{table:tbl4} corresponds to the marginalized $95\%$ 1D confidence intervals for the parameters. As our aim is not to show or solve the $H_{0}$ tension of the low redshift data with the Planck results, in both the combinations of the dataset, we have included the CMB data from the Planck 2018 results. Our aim of this work is to study the behavior of the late time acceleration through a new parametrization which can be useful for larger redshift ranges. In the Table ~\ref{table:tbl4}, 'C', 'B', 'P' and 'H' represent 'CMB', 'BAO', 'Pantheon' and 'H(z)' for OHD data respectively. Here, 'All' means 'CMB+BAO+Pantheon+H+$H_{0}$ combination. Inclusion of $H_{0}$ data shifts the results quite significantly for the cosmological parameters. Mainly the $h$ and $\Omega_{m0}$ parameters change quite significantly. Pade parameters remain almost the same, especially, the parameter $P_{0}$.

One interesting thing is that the values of the Pade parameters are not too high nor too low. This suggests that the Pade series expansion, considered here, is good in nature. Another point to notice here is that the allowed values of $Q_{1}$ and $Q_{2}$ are positive. This corresponds to the comment we made earlier that singularity issue does not appear when parameter values are not far from the standard cosmological fit.

\section{Derived functions}
To say, our series of parameters used here are unknown to us. So, to get a feeling about the cosmological observables, in this section, we plot some important derived quantities from the results from the above parameter ranges i.e. from the MCMC chains.

\begin{figure}[tbp]
\centering
\includegraphics[width=.45\textwidth]{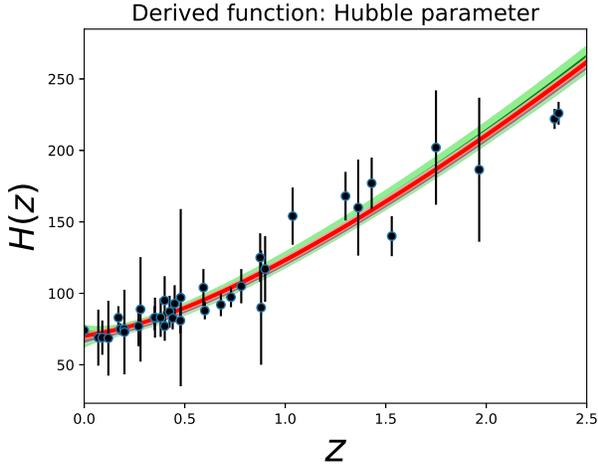}
\caption{\label{fig:derivedH} Derived Hubble parameter. The color codes are the same as in Figure~\ref{fig:triangleplot}. The error bars are from the OHD data for Hubble parameter as a function of redshifts.}
\end{figure}

In Figure~\ref{fig:derivedH}, we have plotted the derived Hubble parameter at 1$\sigma$ and 2$\sigma$ confidence intervals. The color codes are the same as in Figure~\ref{fig:triangleplot}. The error bars are from the OHD data for the Hubble parameter as a function of redshifts. We can see that different combinations considered here can give tight constraints on the evolution of the Hubble parameter.

\begin{figure}[tbp]
\centering
\includegraphics[width=.45\textwidth]{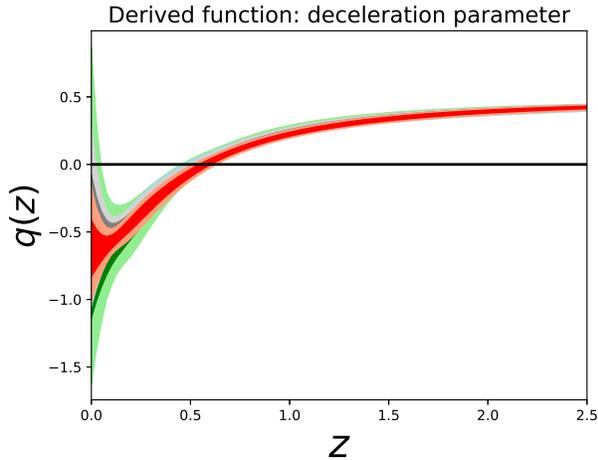}
\caption{\label{fig:derivedq} Derived deceleration parameter (using Eq.~\eqref{eq:qofz}). The color codes are same as in Figure~\ref{fig:triangleplot}. The horizontal black line corresponds to $q=0$ i.e. no acceleration.}
\end{figure}

In Figure~\ref{fig:derivedq}, we have plotted the derived deceleration parameter ($q$) at 1$\sigma$ and 2$\sigma$ confidence intervals. The deceleration parameter ($q$) is defined as

\begin{equation}
q (z) = \frac{1+z}{E(z)} \dfrac{dE(z)}{dz}-1.
\label{eq:qofz}
\end{equation}

\noindent
The deceleration parameter corresponds to the first derivative of the Hubble parameter (or second derivative of the scale factor). This is why it represents the amount of acceleration/deceleration of the Universe. The horizontal black line in Figure~\ref{fig:derivedq} represents no acceleration ($q(z)=0$). from Figure~\ref{fig:derivedq}, we can see that earlier (at matter dominated era) the expansion of the Universe was decelerating. At late time, it becomes accelerating. And the transition from the deceleration to the acceleration occurs near $z=0.5$.

We can derive the equation of state of the dark energy from the Hubble parameter and the relation is given by 

\begin{equation}
w_{DE} (z) = \frac{ \frac{2}{3} (1+z) E(z) \dfrac{dE(z)}{dz} - E^{2}(z)}{E^{2}(z)-\Omega_{m0} (1+z)^3}.
\label{eq:wDE}
\end{equation}

\begin{figure}[tbp]
\centering
\includegraphics[width=.45\textwidth]{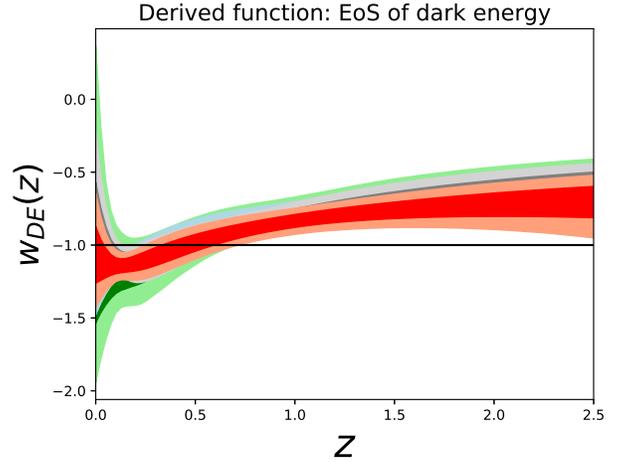}
\caption{\label{fig:derivedwDE} Derived equation of state of the dark energy (using Eq.~\eqref{eq:wDE}). The color codes are same as in Figure~\ref{fig:triangleplot}. The horizontal black line corresponds to the $\Lambda$CDM value. The lower and upper regions of this horizontal black line correspond to the phantom and non-phantom regions respectively.}
\end{figure}

\noindent
In Figure~\ref{fig:derivedwDE}, we have plotted the derived equation of state of the dark energy $w_{DE}(z)$ at 1$\sigma$ and 2$\sigma$ confidence intervals using Eq.~\eqref{eq:wDE}. The color combinations are the same as in Figure~\ref{fig:triangleplot}. First of all, from Figure~\ref{fig:derivedwDE}, we can see that the phantom crossing occurs at around redshift $z=0.3-0.5$. At lower redshifts ($z<0.5$), the allowed values of $w_{DE}(z)$ is phantom at more than 1$\sigma$ confidence level. We can also see that the $\Lambda$CDM model is 1$\sigma$ to 2$\sigma$ away except the phantom crossing regions ($z=0.3-0.6$). And it is also clear that the dynamical dark energy models are preferable at lower redshifts.

\section{Comparison to the CPL parametrization}

In this section, we compare best-fit results (obtained from the real data as mentioned in section-\ref{sec-dataanalysis}) of our series with the CPL parametrization. Here, we do a similar kind of comparison as in section-\ref{sec-fiducialfitting}, but here we do not consider any arbitrary fiducial model. Instead, we compute the best fit parameter values for these two models (our series and the CPL parametrization) using real data, mentioned in  section-\ref{sec-dataanalysis}. We have already computed the best fit parameter values of our series in section-\ref{sec-dataanalysis}. here, we do the same analysis for the CPL parametrization too. By doing the same data analysis (as in section-\ref{sec-dataanalysis}), we find best fit CPL parameter values, listed in Table ~\ref{table:tbl5}.

\begin{table}[tbp]
    \begin{center}
\begin{tabular} {l l l l c c c c}

\hline 
{         } & \text{C+B}  & \text{C+B+P}   & \text{C+B+P+H}   & \text{All}    \\
\hline                                                     \\

{\boldmath$\Omega_{m0}    $} & $0.310^{+0.034}_{-0.034}   $ & $0.310^{+0.017}_{-0.016}   $ & $0.315^{+0.015}_{-0.014}   $ & $0.301^{+0.012}_{-0.011}   $\\ \\

{\boldmath$h              $} & $0.690^{+0.057}_{-0.052}   $ & $0.688^{+0.028}_{-0.027}   $ & $0.679^{+0.022}_{-0.022}   $ & $0.704^{+0.019}_{-0.018}   $\\ \\

{\boldmath$w_{0}          $} & $-1.16^{+0.31}_{-0.28}     $ & $-1.15^{+0.15}_{-0.14}     $ & $-1.11^{+0.15}_{-0.14}     $ & $-1.14^{+0.15}_{-0.15}     $\\ \\

{\boldmath$w_{a}          $} & $0.59^{+0.72}_{-0.78}      $ & $0.55^{+0.49}_{-0.52}      $ & $0.50^{+0.47}_{-0.54}      $ & $0.43^{+0.53}_{-0.57}      $\\ \\
\hline
\end{tabular}
    \end{center}
\caption{Marginalized $95\%$ 1D confidence intervals of the CPL parameters and the cosmological parameters from the different combinations of data mentioned in section-\ref{sec-dataanalysis}.}
\label{table:tbl5}
\end{table}

\begin{figure}[tbp]
\centering
\includegraphics[width=.45\textwidth]{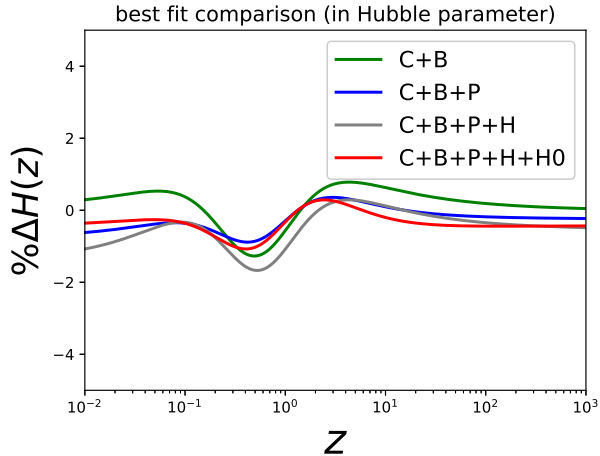}
\caption{\label{fig:Hcmp} Percentage deviations in the best fit evolution of the Hubble parameter in our series compared to the same in the CPL parametrization. The color codes are same as in Figure~\ref{fig:triangleplot}.}
\end{figure}

\noindent
Now, we compute derived Hubble parameter ($H(z)$) for both these models with their respective best fit parameter values. Then, we compare these two $ H (z)$ in Figure~\ref{fig:Hcmp}. We can see that the best fit result of our series matches to the best fit CPL parametrization at around $1\%$ level.

\begin{figure}[tbp]
\centering
\includegraphics[width=.45\textwidth]{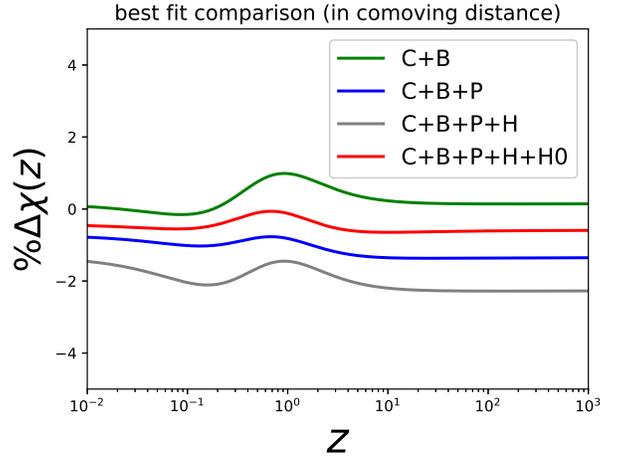}
\caption{\label{fig:comvcmp} Percentage deviations in the best fit evolution of the comoving distance in our series compared to the same in the CPL parametrization. The color codes are same as in Figure~\ref{fig:triangleplot}.}
\end{figure}

\noindent
In Figure~\ref{fig:comvcmp}, we compare the derived comoving distance (line of sight) between both these best fit values. here, we can also see the same result. The results are always at around $1\%$ level agreement at all redshifts (as high as $z=1000$).

In summary, our series method is as good as the CPL parametrization. On the other hand, the Taylor series or the normal Pade series are restricted to the lower redshifts only.

Note that, in this section, we could not compute best-fit results for the Taylor series or the normal Pade series unlike in section-\ref{sec-fiducialfitting}. This is because we can not use these series for higher redshifts. For example, in the CMB likelihood, we need information at higher redshift. These series can not be used at higher redshifts except one considers these series at lower redshifts and replace these series by any standard cosmological model at higher redshifts. For example, in \citep{Dutta:2018vmq}, authors use normal Pade series at lower redshifts and the $\Lambda$CDM model at higher redshifts to do the data analysis.

So, our method always has an advantage that we can use our series both at lower redshifts and at higher redshifts.

\section{Conclusion}

We have considered a model independent parametrization to the comoving distance and consequently to the Hubble parameter. This parameterization can be used up to any higher redshift (up to matter dominated era only) with sub-percentage accuracy. We have neglected the curvature term.

Our methodology can be easily extended for the inclusion of radiation and curvature terms. We shall extend this in the near future.

Although to study the background cosmology for the late time cosmic acceleration, one may not need comoving distance or Hubble parameter up to higher redshift (considering the low redshift data only), but at perturbation level, one must need the values of the Hubble parameter at higher redshift. This is because of the need for the initial condition of the perturbation. So, our parametrization for the Hubble parameter can still be used in the perturbation. Note that this is only valid if we consider initial conditions at early matter dominated era only. If one considers initial conditions at radiation dominated era or even at higher redshift, our parametrization will not be valid.

We have first parametrized the comoving distance and from this, we have consequently parametrized the Hubble parameter. In this way, both quantities have analytical forms. This is an advantage since both the parameters are involved in the background cosmological data.

To study the perturbation equations, one must need the cosmological parameter $\Omega_{m0}$ explicitly. So, any parametrization to the Hubble parameter which does not have $\Omega_{m0}$ as an explicit parameter (due to the direct parametrization of the Hubble parameter), needs to introduce this extra parameter to study the perturbation equations. We shall use our parametrization at the perturbation level in the extension of this work. Since, we have $\Omega_{m0}$ parameter explicitly, we do not need to introduce it in the perturbation equations.

Next, we use the recent low redshift observations (mentioned in the main text) including Planck 2018 results (for TT,TE,EE+lowl+lowE for the base $\Lambda$CDM model) to put constraints on the Pade parameters as well as on the cosmological parameters.

We find that the phantom crossing is allowed for all the combinations of datasets and it occurs at around redshift $z=0.3-0.5$.

Our investigation shows that for low redshift, $\Lambda$CDM model is almost 1$\sigma$ to 2$\sigma$ away except at redshift ($z=0.3-0.6$), where phantom crossing occurs.

Our results also show that dynamical dark energy is preferred at lower redshifts.

We also find that at lower redshifts ($z<0.5$), phantom models are allowed at almost 1$\sigma$ confidence level.


\end{document}